\documentclass{article}
\usepackage{amsmath}
\usepackage{graphicx}
\usepackage{tikz}
\usetikzlibrary{graphs}
\usepackage{verbatim}
\begin{document}

\title{An analytical solution of the gyrokinetic equation for the calculation of neoclassical effects}
\author{Andrea Casolari}
\date{\today}
\maketitle

\section{Introduction}
The purpose of this document is to find an analytical solution for the gyrokinetic equation under specific, simplificative hypotheses. The case I am considering is that of a collisional plasma in the presence of a chain of magnetic islands. The presence of the magnetic islands causes the onset of perturbative fields, in particular an electrostatic field, with a gradient length-scale comparable with the island's width. When the island's width $w$ becomes comparable with the ion Larmor radius $\rho_i$, the drift-kinetic equation is inadequate to treat the transport and the calculation of the neoclassical effects. Nevertheless, I'm going to solve the equation with the methods described by S. P. Hirshman and D. J. Sigmar in the review paper "Neoclassical transport of impurities in tokamak plasmas" \cite{1}, which was developed to solve the drift-kinetic equation in different regimes of collisionality. I'm going to remind first the drift-kinetic theory, which was largely used to study classical and neoclassical transport in magnetized plasmas. Then I'm moving to the gyrokinetic theory, which brings to a more complicated kinetic equation, and I'm going to solve it by applying the approach used previously by Hirshman \& Sigmar.

\section{Drift-kinetic equation}
The purpose of these first sections is to remind the method which was used by several authors, among which Hirshman \& Sigmar \cite{1} and Helander \& Sigmar \cite{2}, to study neoclassical transport We use the coordinate system $w=(\boldsymbol R,E,\mu,\theta)$, where
\begin{equation}
E=\frac{mv^2}{2}+Ze\phi;\hspace{10mm}\mu=\frac{mv_{\perp}^2}{2B}
\end{equation}
and $\theta$ is the gyrophase, the kinetic equation becomes
\begin{equation}
\frac{\partial f}{\partial t}+ \dot{\boldsymbol R}\cdot \nabla f+\dot{E}\frac{\partial f}{\partial E}+\dot{\mu}\frac{\partial f}{\partial \mu}+\dot{\theta}\frac{\partial f}{\partial \theta}=C(f)
\label{1}
\end{equation}
The convective therm can be written as $\dot{\boldsymbol R}\cdot \nabla f=v_{\parallel}\nabla_{\parallel}f+\boldsymbol v_d\cdot \nabla f$, where $\boldsymbol v_d$ is the drift velocity of particles:
\begin{equation}
\boldsymbol v_d=\frac{1}{\Omega}\hat{b}\wedge\left[\left(\mu+\frac{v_{\parallel}^2}{B}\right)\nabla B+\frac{e}{m}\nabla\phi\right]=-v_{\parallel}\hat{b}\wedge\nabla\left(\frac{v_{\parallel}}{\Omega}\right)
\label{2}
\end{equation}
Eq.\ref{2} comes from the fact that $v_{\parallel}=\sqrt{2(E-\mu B-e\phi/m)}$. To come to the drift-kinetic equation, we have to use the finite-Larmor-radius ordering $\delta=\rho/L$ and the hypothesis of strongly-magnetized plasma $\Delta=\nu/\Omega$. In particular:
\begin{equation}
v_{\parallel}\nabla_{\parallel}f\approx\frac{\rho}{L}\Omega f;\hspace{10mm}C(f)\approx\nu(f_0-f)
\end{equation}
whence
\begin{equation}
\frac{v_{\parallel}\nabla_{\parallel}f}{C(f)}\approx \frac{\rho}{L}\frac{\Omega}{\nu}
\end{equation}
If the mean-free-path $\lambda$ is comparable with the length-scale $L$, $\delta\approx \Delta$ follows, and thus the two previous terms are of the same order. Then $\dot{\theta}=\Omega$, so that the term with the $\theta$-derivative is larger than the others by a factor $\delta^{-1}$. To order $\delta^{-1}$ we find that $f_0$ is gyrotropic. To the next order ($\delta^0$) we find
\begin{equation}
v_{\parallel}\nabla_{\parallel}f_0+\Omega\frac{\partial f_1}{\partial \theta}=C(f_0)
\end{equation}
Taking the average over the gyrophase, the term with the $\theta$-derivative becomes zero and we are left with
\begin{equation}
v_{\parallel}\nabla_{\parallel}f_0=C(f_0)
\label{3}
\end{equation}
The solution of Eq.\ref{3} is a Maxwellian which is constant on the flux-surfaces. In fact, if we multiply both members of Eq.\ref{3} by $\log f_0$, we integrate over the velocities and we take the flux-surface average, we find:
\begin{equation}
\left\langle\int \log f_0 C(f_0)d^3v \right\rangle=0
\label{logf_C(f)}
\end{equation}
for each particle species. Ion-electron collisions are weak, $C_{ie}/C_{ii}\approx(m_e/m_i)^{1/2}\ll 1$. According to Boltzman's H-theorem:
\begin{equation}
\int \log f_{0i} C(f_{0i})d^3v\leq 0
\label{H_theorem}
\end{equation}
with the equality holding only if $f_{0i}$ is a Maxwellian. To the next order ($\delta$):
\begin{equation}
v_{\parallel}\nabla_{\parallel}f_1+\boldsymbol v_d\cdot \nabla f_M+ZeE_{\parallel}^{(A)}\frac{\partial f_M}{\partial E}=C(f_1)
\label{4}
\end{equation}
where $E_{\parallel}^{(A)}$ is the induced electric field. Using Eq.\ref{2} and the fact that $\boldsymbol v_d\cdot \nabla f$ represents the flux across the flux-surfaces, Eq.\ref{4} becomes
\begin{equation}
v_{\parallel}\nabla_{\parallel}\left(f_1+\frac{Iv_{\parallel}}{\Omega}\frac{\partial f_M}{\partial \psi}\right)-\frac{ZeE_{\parallel}^{(A)}}{T}v_{\parallel}f_M=C(f_1)
\label{5}
\end{equation}
where $I=RB_{\varphi}$. To get rid of the term with the electric field, we introduce the Spitzer function $f_s$, satisfying the equation
\begin{equation}
C(f_s)=-\frac{ZeE_{\parallel}^{(A)}}{T}v_{\parallel}f_M
\label{6}
\end{equation}
so that Eq.\ref{5} becomes
\begin{equation}
v_{\parallel}\nabla_{\parallel}(f_1-F)=C(f_1-f_s)
\label{7}
\end{equation}
where $F$ has been introduced:
\begin{equation}
F=-\frac{Iv_{\parallel}}{\Omega}\frac{\partial f_M}{\partial \psi}=-\frac{Iv_{\parallel}}{\Omega}\left[\frac{\partial \log n}{\partial \psi}+\frac{Ze}{T}\frac{\partial \phi}{\partial \psi}+\left(\frac{mv^2}{2T}-\frac{3}{2}\right)\frac{\partial \log T}{\partial \psi}\right]f_M
\label{8}
\end{equation}

\section{Banana regime}
Eq.\ref{7} is the drift-kinetic equation, which is the starting point for studying the neoclassical effects. I'm interested here in the low collisionality regime, the so called banana regime. It is usefull, in order to solve Eq.\ref{7}, to expand $f_1$ in a power series respect to the collisionality parameter.
\begin{equation}
\nu^*=\frac{\nu_{eff}}{\omega_b}\ll 1
\end{equation}
where $\nu_{eff}$ is the effective collision frequency, which takes into account the fraction of trapped particles, and $\omega_b$ is the bounce-frequency of particles in the banana orbits. I use the following notation for this secondary expansion: $f_1=f_1^{(0)}+f_1^{(1)}+\cdots$. The parameter $\nu^*$ is much smaller than one in the banana regime, but not as much as $\delta$, which thus remains the primary expansion parameter. We will obtain in the next calculations quantities of order $\delta\nu^*$, which are always much larger than $\delta^2$. We will neglect all the terms of order $\delta^2$ and beyond.
To lowest orders:
\begin{equation}\begin{split}
&v_{\parallel}\nabla_{\parallel}(f_1^{(0)}-F)=0\\&
v_{\parallel}\nabla_{\parallel}f_1^{(1)}=C(f_1^{(0)}-f_s)
\label{9}
\end{split}\end{equation}
The first one of Eq.\ref{9} has the solution $f_1^{(0)}=F+g$, where $g$ is such that $\nabla_{\parallel}g=0$. If we multiply the second of Eq.\ref{9} by $B/v_{\parallel}$ and we take the flux-surface average:
\begin{equation}
\left\langle\frac{B}{v_{\parallel}}C(g+F-f_s)\right\rangle=0
\label{10}
\end{equation}
The function $g$ is zero for trapped particles for parity reasons, so we just have to solve Eq.\ref{10} for the circulating particles. In fact, $g$ must be even respect to $v_{\parallel}$ for the trapped particles, because this is is true in the reflection points $\theta=\pm \theta_b$, but $\partial g/\partial \theta=0$ because $\nabla_{\parallel}g=0$ and $\partial g/\partial \varphi=0$ due to axisymmetry. We can rewrite the second of Eq.\ref{9} in this way:
\begin{equation} 
\frac{\partial f_1^{(1)}}{\partial \theta}=\frac{B}{\sigma|v_{\parallel}|\boldsymbol B\cdot \nabla\theta}C(g+F-f_s)
\end{equation}
Integrating between the reflection points we obtain
\begin{equation} 
\int_{-\theta_b}^{\theta_b}\frac{d\theta}{\boldsymbol B\cdot \nabla\theta}\frac{B}{|v_{\parallel}|}C(g(\sigma=+1)+g(\sigma=-1))=0
\label{odd_condition}
\end{equation}
where the terms containing $F$ and $f_s$ are zero because these two functions are odd respect to $v_{\parallel}$. For condition \ref{odd_condition} to hold, $g$ must be odd respect to $v_{\parallel}$, but we saw before that it must be even, so $g$ must be zero identically in the trapped region.

\section{Transport for the ions}
The ion-ion collision term is much larger than the electron-ion one, which can be neglected. I use the following model for the collision operator, which is well suited to deal with self-collisions:
\begin{equation}
C_{ii}(f_i)=\nu_D^{ii}(v)\left(\mathcal{L}(f_{i1})+\frac{m_iv_{\parallel}u_i}{T_i}f_{Mi}\right)
\label{11}
\end{equation}
$\mathcal{L}$ is the Lorentz operator, representing the pitch-angle contribution to the scattering, which can be written in terms of the pitch-angle parameter $\lambda=v_{\perp}^2B_0/(v^2B)$:
\begin{equation}
\mathcal{L}=\frac{2hv_{\parallel}}{v^2}\frac{\partial}{\partial \lambda}\lambda v_{\parallel}\frac{\partial}{\partial \lambda}
\label{12}
\end{equation}
where $h=B_0/B$ is the toroidal metric coefficient. The velocity $u_i$ is needed for momentum conservation in the collisions:
\begin{equation}
u_i=\int v_{\parallel}\nu_D^{ii}(v) f_i d^3v\Big/\int \nu_D^{ii}(v)\frac{m_iv^2}{3T_i}f_{Mi}d^3v
\label{13}
\end{equation}
The velocity-dependent collision frequency $\nu_D^{ii}(v)$ can be expressed in terms of the error function and of its derivative, and it is an even function of $v$. The collision operator Eq.\ref{11} automatically conserves the particles number and the energy:
\begin{equation}\begin{split}
&\int d^3v C(f)=0\\&
\int d^3v \frac{mv^2}{2}C(f)=0
\end{split}\end{equation}
Using Eq.\ref{13}, Eq.\ref{10} becomes
\begin{equation}
\left\langle \frac{B}{v_{\parallel}}\left[\mathcal{L}(g_i+F_i)+\frac{m_iv_{\parallel}u_i}{T_i}f_{Mi}\right] \right\rangle=0
\label{14}
\end{equation}
where I have neglected the Spitzer function because $f_{si}\ll g_i$ for the ions. Using the property of the Lorentz operator $\mathcal{L}(v_{\parallel})=-v_{\parallel}$ (which can be easily verified) and $\mathcal{L}(F_i)=-F_i$ (which follows from $\mathcal{L}(F_M)=0$, because the Maxwellian is isotropic respect to velocity), Eq.\ref{14} becomes
\begin{equation}
\frac{\partial}{\partial \lambda}\lambda \left\langle v_{\parallel}\right\rangle\frac{\partial g_i}{\partial \lambda}=-\frac{v^2}{2}\left(\frac{I}{h\Omega_i}\frac{\partial \log f_{Mi}}{\partial \psi}+\left\langle \frac{u_i}{h} \right\rangle\frac{m_i}{T_i}\right)f_{Mi}
\label{15}
\end{equation}
The function $g_i$ is zero for trapped particles, so we have to solve it only for circulating particles, such that $0\leq \lambda \leq \lambda_c$, with $\lambda_c=B_0/B_{max}$. If we impose that $g_i$ be continuous, we get the solution
\begin{equation}
g_i=H(\lambda_c-\lambda)V_{\parallel}s_iF_{Mi}
\label{16}
\end{equation}
where $V_{\parallel}$ is so defined:
\begin{equation}
V_{\parallel}=\frac{\sigma v}{2}\int_{\lambda}^{\lambda_c}\frac{d\lambda'}{\left\langle \sqrt{1-\lambda'/h(\theta)} \right\rangle}
\label{v_par_maiusc}
\end{equation}
$H(\lambda_c-\lambda)$ is the Heaviside step function
\begin{equation}
H(\lambda_c-\lambda)=\left\{\begin{array}{ll}
1\hspace{10mm}0\leq\lambda\leq\lambda_c\\
0\hspace{13mm}\lambda\geq\lambda_c
\end{array}
\right.
\label{step_function}
\end{equation}
In the large-aspect-ratio limit $\epsilon \rightarrow 0$, we find $V_{\parallel}\rightarrow v_{\parallel}$.

\section{Complete solution}
The complete distribution function $f_{i1}=g_i+F_i$ is
\begin{equation}
f_{i1}=-\frac{I}{h\Omega_i}(hv_{\parallel}-HV_{\parallel})\frac{\partial f_{Mi}}{\partial \psi}+\frac{m_iHV_{\parallel}}{T_i}\left\langle \frac{u_i}{h} \right\rangle f_{Mi}
\label{17}
\end{equation}
Intorducing the average over the velocities so defined
\begin{equation}
\{F(v)\}=\int F\frac{mv^2}{nT}f_Md^3v=\frac{8}{3\sqrt{\pi}}\int_0^{\infty}F(x)e^{-x^2}x^4dx
\label{18}
\end{equation}
Eq.\ref{12} can be written in this way:
\begin{equation}
\{\nu_D^{ii}\}\left\langle \frac{u_i}{h} \right\rangle=\left\langle \frac{1}{hn_i}\int\nu_D^{ii}v_{\parallel}f_{i1}d^3v \right\rangle
\label{19}
\end{equation}
For every function $F(v)$, the following property holds
\begin{equation}
\left\langle \int F(v)\frac{mv_{\parallel} HV_{\parallel}}{hT}f_Md^3v \right\rangle=(1-f_t)n\{F\}
\label{19+1}
\end{equation}
where the fraction of trapped particles $f_t$ was introduced, whence
\begin{equation}
\left\langle \int F(v)\frac{mv_{\parallel} (hv_{\parallel}-HV_{\parallel})}{hT}f_Md^3v \right\rangle=f_tn\{F\}
\label{19+2}
\end{equation}
Substituing Eq.\ref{17} into Eq.\ref{19}, we obtain
\begin{equation}
\left\langle \frac{u_i}{h} \right\rangle=-\frac{IT_i}{hm_i\Omega_i}\frac{\{\nu_D^{ii}\partial \log f_{Mi}/\partial \psi\}}{\{n\nu_D^{ii}\}}\approx -\frac{IT_i}{hm_i\Omega_i}\left(\frac{d\log P_i}{d\psi}+\frac{Ze}{T_i}\frac{d\phi}{d\psi}-1.173\frac{d\log T_i}{d\psi}\right)
\label{proprieta_media}
\end{equation}
where $-1.173=\{\nu_D^{ii}(x^2-5/2)\}/\{\nu_D^{ii}\}$. The complete distribution function Eq.\ref{17} becomes
\begin{equation}
f_{i1}=-\frac{Iv_{\parallel}}{\Omega_i}\frac{\partial f_{Mi}}{\partial \psi}+\frac{IHV_{\parallel}}{h\Omega_i}\left(\frac{m_iv^2}{2T_i}-1.33\right)\frac{d\log T_i}{d\psi}f_{Mi}
\label{20}
\end{equation}
The poloidal component of the plasma rotation velocity is all contained in the term proportional to the temperature gradient:
\begin{equation}
K_i=1.17f_c\frac{n_iI}{m_i\Omega_iB_0}\frac{dT_i}{d\psi}
\end{equation}
where $f_c=1-f_t$ is the fraction of circulating particles. In the large-aspect-ratio limit $\epsilon\rightarrow 0$, the parallel velocity of the plasma becomes
\begin{equation}
U_{\parallel i}=\int d^3v v_{\parallel}f_1^{(0)}=-\frac{IT_i}{m_i\Omega}\left(\frac{d\log P_i}{d\psi}+\frac{Ze}{T_i}\frac{d\phi}{d\psi}-1.17\frac{d\log T_i}{d\psi}\right)
\label{parallel_flow_short}
\end{equation}
The neoclassical poloidal flow damping is related to the radial particle flux in the banana-plateau regime by the following flux-friction relation:
\begin{equation}
\left\langle \Gamma\cdot\nabla \psi\right\rangle^{BP}\equiv -I\frac{\left\langle \boldsymbol B\cdot \nabla \cdot \boldsymbol\pi\right\rangle}{Ze\left\langle B^2\right\rangle}
\label{flux_stress}
\end{equation}
The right-hand side in turn is related to the parallel component of the friction force and to the parallel induced electric field:
\begin{equation}
\left\langle \boldsymbol B\cdot \nabla \cdot \boldsymbol\pi\right\rangle=\left\langle B(F_{\parallel}+nZeE_{\parallel}^{(A)})\right\rangle
\label{stress_friction}
\end{equation}
These relations enable us to compute the poloidal flow damping both as an effect of collisions or as  consequence of pressure anisotropy (which is contained in the tensor $\boldsymbol \pi$). Using Eq.\ref{stress_friction}, together with the definition
\begin{equation}
F_{\parallel}\equiv\int mv_{\parallel}C(f_1)d^3v
\end{equation}
we can compute $\left\langle \boldsymbol B\cdot\nabla\cdot\boldsymbol \pi_i\right\rangle$ with the distribution function \ref{f1_shorter}. First of all we introduce the Spitzer function $f_{is}$ which solves Eq.\ref{6}, so that Eq.\ref{stress_friction} becomes
\begin{equation}
\left\langle \boldsymbol B\cdot\nabla\cdot\boldsymbol \pi_i\right\rangle=\left\langle B\int mv_{\parallel}C(f_{1i}^{(0)}-f_{is})d^3\right\rangle
\label{stress_friction_simp}
\end{equation}
For the ions, the Spitzer function can be neglected because the ion mass is much larger than the electrons, so that the acceleration caused by the electric field is much smaller for the ions than for the electrons. Using Eq.\ref{11} for the collision operator in Eq.\ref{stress_friction_simp}:
\begin{equation}
\left\langle \boldsymbol B\cdot\nabla\cdot\boldsymbol \pi_i\right\rangle=\left\langle B\int mv_{\parallel}\nu_D^{ii}\left(\mathcal{L}(f_1^{(0)})+\frac{mv_{\parallel}U_{\parallel i}}{T}F_M\right)d^3v\right\rangle
\label{formula_85}
\end{equation}
Using the properties $\mathcal{L}(v_{\parallel})=-v_{\parallel}$ and $\mathcal{L}(F_M)=0$, we find out that $\mathcal{L}(f_1^{(0)})=-f_1^{(0)}$. I used $U_{\parallel i}$ instead of $u_i$ in the collision operator because Eq.\ref{13} is basically the definition of the parallel flow velocity in the large-aspect-ratio limit. Performing the calculations by using the solution Eq.\ref{f1_shorter}, we find
\begin{equation}
\left\langle \boldsymbol B\cdot\nabla\cdot\boldsymbol \pi_i\right\rangle=0.53\nu_{ii}B1.17\frac{InT_i}{\Omega_i}\frac{d\log T_i}{d\psi}
\label{grad_temp}
\end{equation}
Eq.\ref{grad_temp} can be written as
\begin{equation}
\left\langle \boldsymbol B\cdot\nabla\cdot\boldsymbol \pi_i\right\rangle=B\mu_{01}m_in\nu_{ii}V_{i\theta}
\label{pol-flow-dam_definitive}
\end{equation}
where $\mu_{01}\approx 0.53$ and $V_{i\theta}$ is the neoclassical poloidal velocity:
\begin{equation}
V_{i\theta}=1.17\frac{IT_i}{m_i\Omega_i}\frac{d\log T_i}{d\psi}
\label{poloidal_rotation}
\end{equation}

\newpage

\section{Gyrokinetic ordering}
Moving now to the gyrokinetic case, I try to find the neoclassical results again with the finite-Larmor-radius (FLR) corrections. The purpose of these calculations is to find and expression for the poloidal flow damping which can fit in a system of  four-field gyrofluid equations, which in turn can be applied to study the dynamic of magnetic islands for arbitrary island's width $w$. To deduce the gyrokinetic equation, I will follow in part the thesis "Modelling of Turbulent Particle Transport in Finite-Beta and Multiple Ion Species Plasma in Tokamaks" di Gabor Szepesi \cite{3} and the work by Parra \& Catto "Limitations of gyrokinetics on transport time scales" \cite{4}. In the drift-kinetic theory, the hypothesis is made that the fields vary on scale lengths comparable with the equilibrium scale $L$, so that, when the gyroaverage operation is performed, we can reasonably take the value of the fields in the guiding center positions. In gyrokinetics we assume that the fields can vary on lengths scales comparable to the ion Larmor radius, so that their gyroaverage must be computed explicitly. The main orderings of gyrokinetic theory are the followings:
\begin{equation}\begin{split}
&\left|\frac{A_1}{A_0}\right|\approx \frac{\phi_1}{\phi_0}=\delta_f\ll 1\\&
\frac{\rho_i}{L}=\delta_B\ll 1\\&
\frac{\omega}{\Omega}=\delta_{\omega}\ll 1\\&
k_{\perp}\rho_i\approx 1
\end{split}\end{equation}
In principle $\delta_f$, $\delta_B$ e $\delta_{\omega}$ are independent parameter, but for simplicity I assume they are of the same order. 

\section{Derivation of gyrokinetic equation}
Following the thesis of Gabor Szepesi, I start from Eq.\ref{1}, written as usual in the variables $w=(\boldsymbol R,E,\mu,\theta)$. Initially, in order to be consistent with the equations from the thesis, I will omit the collisional term. We can take the gyroaverage to get rid of the term with the $\theta$-derivative and we pass to the variable $v_{\parallel}$, such that:
\begin{equation}
\dot{E}\frac{\partial f}{\partial E}=\dot{v}_{\parallel}\frac{\partial f}{\partial v_{\parallel}}
\end{equation}
Now $f$ is the gyrokinetic distribution function, so the magnetic moment $\mu$ is constant ($\dot{\mu}=0$). These simplifications being done, Eq.\ref{1} becomes
\begin{equation}
\frac{\partial f}{\partial t}+\dot{\boldsymbol R}\cdot \frac{\partial f}{\partial \boldsymbol R}+\dot{v}_{\parallel} \frac{\partial f}{\partial v_{\parallel}}=0
\label{gy_eq}
\end{equation}
The purpose of gyrokinetic theory is to simplify the kinetic calculations by substituing the particles position with the gyrocenters position, which is associated with the center of the particles orbit in their gyromotion around magnetic field lines. The gyrocenter coordinates can be obtained by doing an appropriate transformation, which first brings the coordinates from the particles position to the position of the center of their gyromotion (guiding center) with the quilibrium fields. In these guiding center coordinates, the magnetic moment is a constant of motion because the distribution function is gyrotropic. When we add the field perturbations, the distribution function is no longer gyrotropic, so we have to perform a new change of coordinates restoring the constancy of the magnetic moment. Instead of performing an additional gyroverage, in modern gyrokinetics (see for example "Foundations of nonlinear gyrokinetic theory" by Brizard \& Hahm \cite{5}) the non-canonical perturbation theory approach is followed, so that the guiding center coordinates are transformed in the gyrocenter coordinates perturbatively by Lie-trasform perturbation theory. In the following I will assume that those trasformations have been done and I will use only the final results.\\
In Eq.\ref{gy_eq}, $\dot{\boldsymbol R}$ and $\dot{v}_{\parallel}$ are
\begin{equation}
\dot{\boldsymbol R}=v_{\parallel}\left(1+\frac{\bar{B}_{1\perp}}{B}\right)\hat{b}+\boldsymbol{v}_d
\end{equation}
where $\bar{B}_{1\perp}$ is the perturbation to the magnetic field perpendicular to the diretcion of $B_0$, while $\boldsymbol{v}_d$ is the drift velocity of particles, in the presence of the total fields (equilibrium and the perturbations). In this document, by perturbations I mean those fields which result from the onset of a magnetic island, so I'm not dealing with turbulence. The other component needed in Eq.\ref{gy_eq} is
\begin{equation}
\dot{v}_{\parallel}=-\frac{1}{m}\left(1+\frac{\bar{B}_{1\perp}}{B}\right)[Ze\nabla_{\parallel}\bar{\phi}_1+\mu\nabla_{\parallel}(B_0+\bar{B}_{1\parallel})]-\frac{1}{mv_{\parallel}}\boldsymbol{v}_d\cdot[Ze\nabla\bar{\phi}_1+\mu\nabla(B_0+\bar{B}_{1\parallel})]+\frac{Ze}{m}E_{\parallel}^{(A)}
\label{v_par_punto}
\end{equation}
The bar simbol over the perturbed fields represents the gyroaverage operation. Unlike in the corresponding equation in the work by Szepesi, I added in Eq.\ref{v_par_punto} one term proportional to the parallel induced electric field, which is due to the time variation of the magnetic flux during the discharge in tokamaks. This term will be absorbed in a Spitzer function, once collisions are reintroduced. Now the distribution function $f$ is expanded in an equilibrum Maxwellian part plus a small perturbation, orderd with $\delta=\rho/L$: $f=F_M+f_1$. The equilibrium solution is assumed stationary and it is a flux-function, so that
\begin{equation}
\frac{\partial F_M}{\partial t}=0;\hspace{10mm}\nabla_{\parallel}F_M=0
\end{equation}
The following orderings are used: $\partial/\partial t=O(\delta^2 v_{th}/L)$, $f_1/F_M=O(\delta)$, $B_1/B_0=O(\delta)$, $Ze\phi_1/T=O(\delta)$, $v_d/v_{th}=O(\delta)$, $\nabla_{\parallel}=O(1/L)$, $\nabla_{\perp}F_M\approx F_M/L$, $\nabla_{\perp}f_1\approx \delta F_M/\rho=F_M/L$, $\nabla_{\perp}B_1\approx \delta B_0/\rho=B_0/L$. Unlike in drift-kinetic theory, in gyrokinetics the perpendicular gradients of perturbations are $O(1)$ because the perturbations vary on length scales comparable with $\rho_i$, while the equilibrium fields vary on the length scale $L\gg\rho_i$. With these orderings, the gyrokinetic equation to order $\delta$ becomes
\begin{equation}\begin{split}
v_{\parallel}\nabla_{\parallel}f_1+&\boldsymbol{v}_d\cdot\nabla(f_1+F_M)-\frac{\mu}{m}\nabla_{\parallel}B_0\frac{\partial f_1}{\partial v_{\parallel}}-\frac{1}{m}\left[Ze\nabla_{\parallel}\bar{\phi}_1+\mu\left(\nabla_{\parallel}\bar{B}_{1\parallel}+\frac{\bar{B}_{1\perp}}{B}\nabla_{\parallel}B_0\right)\right]\frac{\partial F_M}{\partial v_{\parallel}}-\\&-\frac{1}{mv_{\parallel}}\boldsymbol{v}_d\cdot[Ze\nabla\bar{\phi}_1+\mu\nabla(B_0+\bar{B}_{1\parallel})]\frac{\partial F_M}{\partial v_{\parallel}}+\frac{Ze}{m}E_{\parallel}^{(A)}\frac{\partial F_M}{\partial v_{\parallel}}=0
\end{split}\end{equation}
Making the further assumption that $\nabla B_0\ll \nabla \phi_1,\nabla B_{1\parallel}$, meaning that the equilibrium magnetic field is almost uniform:
\begin{equation}
v_{\parallel}\nabla_{\parallel}f_1+\boldsymbol{v}_d\cdot\nabla(f_1+F_M)+\frac{F_M}{T}v_{\parallel}[Ze\nabla_{\parallel}\bar{\phi}_1-ZeE_{\parallel}^{(A)}+\mu\nabla_{\parallel}\bar{B}_{1\parallel}]+\frac{F_M}{T}\boldsymbol{v}_d\cdot[Ze\nabla\bar{\phi}_1+\mu\nabla\bar{B}_{1\parallel}]=0
\end{equation}
Using Eq.\ref{2} to express $\boldsymbol{v}_d$ in terms of $v_{\parallel}$, the gyrokinetic equation becomes
\begin{equation}
v_{\parallel}\nabla_{\parallel}\left[f_1+\frac{Iv_{\parallel}}{\Omega}\frac{\partial}{\partial \psi}(f_1+F_M)\right]=-\frac{F_M}{T}v_{\parallel}\left\{[Ze\nabla_{\parallel}\bar{\phi}_1-ZeE_{\parallel}^{(A)}+\mu\nabla_{\parallel}\bar{B}_{1\parallel}]+\nabla_{\parallel}\left[\frac{Iv_{\parallel}}{\Omega}\left(Ze\frac{\partial\bar{\phi}_1}{\partial \psi}+\mu\frac{\partial\bar{B}_{\parallel 1}}{\partial \psi}\right)\right]\right\}
\label{girocinetica}
\end{equation}

\section{Collisional gyrokinetic equation}
Gyrokinetic theory is usually used to study turbulent transport, which is typically much larger than the  collisional one. For this reason the gyrokinetic equation deduced in the previous section didn't have the collisional term on the right-hand side. However, to find the results from the drift-kinetic theory again, I have to reintroduce the the collision operator.\\
The drift kinetic equation Eq.\ref{5} without the electric field contribution is
\begin{equation}
v_{\parallel}\nabla_{\parallel}\left[f_1+\frac{Iv_{\parallel}}{\Omega}\frac{\partial F_M}{\partial\psi}\right]=C(f_1)
\end{equation}
When including the collision operator, Eq.\ref{girocinetica} becomes:
\begin{equation}
v_{\parallel}\nabla_{\parallel}\left[f_1+\frac{Iv_{\parallel}}{\Omega}\frac{\partial}{\partial \psi}(f_1+F_M)\right]=-\frac{F_Mv_{\parallel}}{T}\left\{[Ze\nabla_{\parallel}\bar{\phi}_1+\mu\nabla_{\parallel}\bar{B}_{1\parallel}]+\nabla_{\parallel}\left[\frac{Iv_{\parallel}}{\Omega}\left(Ze\frac{\partial\bar{\phi}_1}{\partial \psi}+\mu\frac{\partial\bar{B}_{\parallel 1}}{\partial \psi}\right)\right]\right\}+C(f_1)
\end{equation}
The contribution from $E_{\parallel}^{(A)}$ was absorbed in a Spitzer function by Eq.\ref{6}. This one is neglected respect to $f_1$ for the reasons explained above. For the fields perturbations caused by the onset of a magnetic island, the leading term is $\bar{B}_{\perp 1}$, so that we can neglect $\bar{B}_{\parallel 1}$. With this simplifications:
\begin{equation}
v_{\parallel}\nabla_{\parallel}\left[f_1+\frac{Iv_{\parallel}}{\Omega}\frac{\partial}{\partial \psi}(f_1+F_M)\right]=-\frac{F_Mv_{\parallel}}{T}\left\{Ze\nabla_{\parallel}\bar{\phi}_1+\nabla_{\parallel}\left[\frac{Iv_{\parallel}}{\Omega}\left(Ze\frac{\partial\bar{\phi}_1}{\partial \psi}\right)\right]\right\}+C(f_1)
\label{nuova_equaz}
\end{equation}
In the low-collisional regime, we can expand $f_1$ in a power series of the collisionality $\nu^*$, so that we can write $f_1=f_1^{(0)}+f_1^{(1)}+\cdots$. To the two lowest orders
\begin{equation}\begin{split}
&\nabla_{\parallel}\left[f_1^{(0)}+\frac{Iv_{\parallel}}{\Omega}\frac{\partial}{\partial \psi}(f_1^{(0)}+F_M)\right]+F_M\frac{Ze}{T}\nabla_{\parallel}\left[\bar{\phi}_1+\frac{Iv_{\parallel}}{\Omega}\frac{\partial \bar{\phi}_1}{\partial \psi}\right]=0\\&
v_{\parallel}\nabla_{\parallel}\left[f_1^{(1)}+\frac{Iv_{\parallel}}{\Omega}\frac{\partial f_1^{(1)}}{\partial \psi}\right]=C(f_1^{(0)})
\label{lowest_orders}
\end{split}\end{equation}
Using the fact that $\nabla_{\parallel}F_M=\nabla_{\parallel}T=0$ (neglecting significant perturbations to the temperature), the lowest order equation becomes
\begin{equation}
\nabla_{\parallel}\left[f_1^{(0)}+\frac{Iv_{\parallel}}{\Omega}\frac{\partial}{\partial \psi}(f_1^{(0)}+F_M)+F_M\frac{Ze}{T}\left(\bar{\phi}_1+\frac{Iv_{\parallel}}{\Omega}\frac{\partial \bar{\phi}_1}{\partial \psi}\right)\right]=0
\end{equation}
By integrating once, we find the following equation for $f_1^{(0)}$:
\begin{equation}
\left(1+\frac{Iv_{\parallel}}{\Omega}\frac{\partial}{\partial \psi}\right)f_1^{(0)}=g-\frac{Iv_{\parallel}}{\Omega}\frac{\partial F_M}{\partial \psi}-F_M\frac{Ze}{T}\left(1+\frac{Iv_{\parallel}}{\Omega}\frac{\partial}{\partial \psi}\right)\bar{\phi}_1
\label{eq_per_f_1^0}
\end{equation}
with $g$ unknown function such that $\nabla_{\parallel}g=0$.

\section{Solution for $f_{1}^{(0)}$}
Eq.\ref{eq_per_f_1^0} can be solved formally, by writing the solution $f_1^{(0)}$ in an integral form. Every time we deal with an equation of this form
\begin{equation}
\left(1+aD\right)f(x)=K(x)
\label{equaz_tipo}
\end{equation}
the particular solution takes the form
\begin{equation}
f(x)=\frac{e^{-x/a}}{a}\int_{x_0}^x e^{y/a}K(y)dy
\end{equation}
Eq.\ref{eq_per_f_1^0} is in the form Eq.\ref{equaz_tipo}, so the solution for $f_1^{(0)}$ becomes
\begin{equation}
f_{1}^{(0)}=c_1(v)e^{-\psi/\psi_s}+\frac{e^{-\psi/\psi_s}}{\psi_s}\int_{\psi_0}^{\psi} e^{\chi/\psi_s}\left[g(\chi)-\psi_s\frac{\partial F_M}{\partial \chi}-Ze\frac{F_M}{T}\left(1+\psi_s\frac{\partial}{\partial\chi}\right)\bar{\phi}_1\right]d\chi
\label{general_solution}
\end{equation}
where $\psi_s=Iv_{\parallel}/\Omega$ has the dimensions of a magnetic flux. In the following, I will assume that the parallel velocity appearing in the definition of $\psi_s$ is $v_{th}$ in order to avoid this velocity dependence, which would bring an additional complication. $c_1(v)e^{-\psi/\psi_s}$ is the solution of the homogeneus equation.

\newpage

\subsection{Limit of large wavelengths}
Coming back to Eq.\ref{nuova_equaz}, which I write again in this form:
\begin{equation}
v_{\parallel}\nabla_{\parallel}\left[(1+\psi_s\partial_{\psi})f_1+\psi_s\partial_{\psi}F_M+Ze\frac{F_M}{T}(1+\psi_s\partial_{\psi})\bar{\phi}_1\right]=C(f_1)
\end{equation}
I consider the limit $\psi_s\partial_{\psi}\ll1$, corresponding to large islands ($w/\rho_i\gg 1$). Eq.\ref{nuova_equaz} becomes
\begin{equation}
v_{\parallel}\nabla_{\parallel}(f_1+\psi_s\partial_{\psi}F_M)+\frac{Ze\nabla_{\parallel}\phi_1}{T}v_{\parallel}F_M=C(f_1)
\label{caso_1}
\end{equation}
I removed the bar symbol over $\phi_1$ because in the limit $\psi_s\partial_{\psi}\ll1$ we can approximate the gyroaverage of the fields with their value in the position of the guiding centers. Another Spitzer function can be introduced, which solves the following equation (totally analogous to Eq.\ref{6}):
\begin{equation}
C(f_s)=\frac{Ze\nabla_{\parallel}\phi_1}{T}v_{\parallel}F_M
\end{equation}
so that Eq.\ref{caso_1} becomes
\begin{equation}
v_{\parallel}\nabla_{\parallel}(f_1+\psi_s\partial_{\psi}F_M)=C(f_1-f_s)
\end{equation}
which is totally analogous to Eq.\ref{7}, where $\psi_s\partial_{\psi}F_M=-F$. From this point we can proceed as before and we find again the results valid in the limit of the drift-kinetic equation.\\

\subsection{General solution}
For an island of arbitrary width, we must face the complete solution Eq.\ref{general_solution}. After integrating by parts and introducing the following function:
\begin{equation}
G=g-\psi_s\partial_{\psi}F_M-Ze\frac{F_M}{T}(1+\psi_s\partial_{\psi})\bar{\phi}_1
\label{G_function}
\end{equation}
Eq.\ref{general_solution} becomes
\begin{equation}
f_1^{(0)}=G(\psi)+e^{-\psi/\psi_s}\left[c_1(v)-\int_{\psi_0}^{\psi}d\chi e^{\chi/\psi_s}\partial_{\chi}G(\chi)\right]
\end{equation}
When writing the collision operator $C$ in the form Eq.\ref{11}, Eq.\ref{10} becomes
\begin{equation}
\left\langle \frac{B}{v_{\parallel}}\left(\mathcal{L}(G(\psi))+e^{-\psi/\psi_s}\left[\mathcal{L}(c_1(v))-\int_{\psi_0}^{\psi}d\chi e^{\chi/\psi_s}\mathcal{L}(\partial_{\chi}G(\chi))\right]+\frac{m_iv_{\parallel}u_i}{T}F_M\right)\right\rangle=0
\label{equaz_awesome_expl}
\end{equation}
where I could bring the Lorentz operator $\mathcal{L}$ inside the integral because it only acts on the velocities. Now we can use the properties of $\mathcal{L}$, which we have used previously, such as $\mathcal{L}(v_{\parallel})=-v_{\parallel}$ and $\mathcal{L}(F_M)=0$. Using these properties and the definition Eq.\ref{G_function}, Eq.\ref{equaz_awesome_expl} becomes
\begin{equation}\begin{split}
&\left\langle\frac{B}{v_{\parallel}}\left\{\mathcal{L}(g)+\psi_s\partial_{\psi}F_M+Ze\frac{F_M}{T}\psi_s\partial_{\psi}\bar{\phi}_1+e^{-\psi/\psi_s}\mathcal{L}(c_1(v))-\right.\right.\\&\left.\left.-e^{-\psi/\psi_s}\int_{\psi_0}^{\psi}d\chi e^{\chi/\psi_s}\left(\mathcal{L}(\partial_{\chi}g)+\psi_s\partial^2_{\chi}F_M+Ze\partial_{\chi}\left(\frac{F_M}{T}\psi_s\partial_{\chi}\bar{\phi}_1\right)\right)+\frac{m_iv_{\parallel}u_i}{T}F_M\right\}\right\rangle=0
\label{equaz_awesome_final}
\end{split}\end{equation}
Gathering the functions which are acted on by $\mathcal{L}$, Eq.\ref{equaz_awesome_final} becomes:
\begin{equation}\begin{split}
&\left\langle B\left\{\frac{2h}{v^2}\frac{\partial}{\partial \lambda}\lambda v_{\parallel}\frac{\partial}{\partial \lambda}\left(g-e^{-\psi/\psi_s}\int_{\psi_0}^{\psi}d\chi e^{\chi/\psi_s}\partial_{\chi}g+e^{-\psi/\psi_s}c_1\right)+\right.\right.\\&\left.\left.+\frac{I}{\Omega}\left[\partial_{\psi}F_M+Ze\frac{F_M}{T}\partial_{\psi}\bar{\phi}_1-e^{-\psi/\psi_s}\int_{\psi_0}^{\psi}d\chi e^{\chi/\psi_s}\left(\partial^2_{\chi}F_M+Ze\partial_{\chi}\left(\frac{F_M}{T}\partial_{\chi}\bar{\phi}_1\right)\right)\right]+\frac{m_iu_i}{T}F_M\right\}\right\rangle=0
\end{split}\end{equation}
We define the auxiliary function $J$:
\begin{equation}
J=g-e^{-\psi/\psi_s}\int_{\psi_0}^{\psi}d\chi e^{\chi/\psi_s}\partial_{\chi}g+e^{-\psi/\psi_s}c_1
\end{equation}
The equation for $J$ is
\small
\begin{equation}\begin{split}
\frac{\partial}{\partial \lambda}\lambda \left\langle v_{\parallel}\right\rangle\frac{\partial}{\partial \lambda}J=-\frac{v^2}{2}\left\{\frac{I}{h\Omega}\left[\partial_{\psi}F_M+Ze\frac{F_M}{T}\partial_{\psi}\bar{\phi}_1-e^{-\psi/\psi_s}\int_{\psi_0}^{\psi}d\chi e^{\chi/\psi_s}\left(\partial^2_{\chi}F_M+Ze\partial_{\chi}\left(\frac{F_M}{T}\partial_{\chi}\bar{\phi}_1\right)\right)\right]+\left\langle \frac{u_i}{h} \right\rangle\frac{m_i}{T}F_M\right\}
\label{equ_per_j}
\end{split}\end{equation}
In the limit of large wavelengths we recover Eq.\ref{15}, which I write here for comparison:
\begin{equation}
\frac{\partial}{\partial \lambda}\lambda \left\langle v_{\parallel}\right\rangle\frac{\partial g}{\partial \lambda}=-\frac{v^2}{2}\left[\frac{I}{h\Omega_i}\partial_{\psi}F_M+\left\langle \frac{u_i}{h} \right\rangle\frac{m_i}{T_i}F_M\right]
\label{equ_per_g}
\end{equation}
The function $g$ must be zero for trapped particles (as we already saw), that is for $\lambda\geq B/B_{max}=\lambda_c$. To solve Eq.\ref{equ_per_g}, we can integrate twice for $0\leq \lambda\leq \lambda_c$, and we find out
\begin{equation}
g=H(\lambda_c-\lambda)\frac{v^2}{2}\int_{\lambda}^{\lambda_c}\frac{d\lambda'}{\left\langle v_{\parallel}(\lambda') \right\rangle}\left[\frac{I}{h\Omega_i}\partial_{\psi}F_M+\left\langle \frac{u_i}{h} \right\rangle\frac{m_i}{T_i}F_M\right]
\end{equation}
where $H$ is the step function Eq.\ref{step_function}. From the form of the equations, we can deduce that $J$ plays the role of $g$ in the drift-kinetic equation, so that the solution of Eq.\ref{equ_per_j} is
\begin{equation}\begin{split}
&J=H(\lambda_c-\lambda)\frac{v^2}{2}\int_{\lambda}^{\lambda_c}\frac{d\lambda'}{\left\langle v_{\parallel}(\lambda') \right\rangle}\left\{\frac{I}{h\Omega}\left[\partial_{\psi}F_M+Ze\frac{F_M}{T}\partial_{\psi}\bar{\phi}_1-\right.\right.\\&\left.\left.-e^{-\psi/\psi_s}\int_{\psi_0}^{\psi}d\chi e^{\chi/\psi_s}\left(\partial^2_{\chi}F_M+Ze\partial_{\chi}\left(\frac{F_M}{T}\partial_{\chi}\bar{\phi}_1\right)\right)\right]+\left\langle \frac{u_i}{h} \right\rangle\frac{m_i}{T}F_M\right\}
\end{split}\end{equation}
Using the function $V_{\parallel}$ defined in Eq.\ref{v_par_maiusc}, we can write $f_1^{(0)}$ as:
\begin{equation}\begin{split}
&f_1^{(0)}=\frac{I}{h\Omega}(HV_{\parallel}-hv_{\parallel})\left[\partial_{\psi}F_M+Ze\frac{F_M}{T}\partial_{\psi}\bar{\phi}_1-e^{-\psi/\psi_s}\int_{\psi_0}^{\psi}d\chi e^{\chi/\psi_s}\left(\partial^2_{\chi}F_M+Ze\partial_{\chi}\left(\frac{F_M}{T}\partial_{\chi}\bar{\phi}_1\right)\right)\right]-\\&-\left[F_M\frac{Ze\bar{\phi}_1}{T}-e^{-\psi/\psi_s}\int_{\psi_0}^{\psi}d\chi e^{\chi/\psi_s}Ze\partial_{\chi}\left(\frac{F_M}{T}\bar{\phi}_1\right)\right]+HV_{\parallel}\left\langle\frac{u_i}{h}\right\rangle\frac{m_i}{T}F_M
\label{f_1^0_completa}
\end{split}\end{equation}
using Eq.\ref{19} and substituing Eq.\ref{f_1^0_completa}:
\begin{equation}\begin{split}
&\left\langle \frac{u_i}{h}\right\rangle-\frac{1}{\{\nu_D^{ii}\}}\left\langle \frac{1}{hn}\int d^3v\nu_D^{ii}v_{\parallel}HV_{\parallel}\frac{m_i}{T}F_M\right\rangle\left\langle\frac{u_i}{h}\right\rangle=\frac{1}{\{\nu_D^{ii}\}}\left\langle\frac{1}{hn}\int d^3v\nu_D^{ii}v_{\parallel}\left\{\frac{I}{h\Omega}(HV_{\parallel}-hv_{\parallel})\left[\partial_{\psi}F_M+Ze\frac{F_M}{T}\partial_{\psi}\bar{\phi}_1-\right.\right.\right.\\&\left.\left.\left.-e^{-\psi/\psi_s}\int_{\psi_0}^{\psi}d\chi e^{\chi/\psi_s}\left(\partial^2_{\chi}F_M+Ze\partial_{\chi}\left(\frac{F_M}{T}\partial_{\chi}\bar{\phi}_1\right)\right)\right]-\left[F_M\frac{Ze\bar{\phi}_1}{T}-e^{-\psi/\psi_s}\int_{\psi_0}^{\psi}d\chi e^{\chi/\psi_s}Ze\partial_{\chi}\left(\frac{F_M}{T}\bar{\phi}_1\right)\right]\right\}\right\rangle
\label{equ_per_ui}
\end{split}\end{equation}
Using the properties Eq.\ref{19+1} and Eq.\ref{19+2}:
\begin{equation}\begin{split}
\left\langle \frac{u_i}{h}\right\rangle=&-\frac{IT}{hm_i\Omega}\frac{1}{\{\nu_D^{ii}\}}\left\{\nu_D^{ii}\left[\frac{\partial_{\psi}F_M}{F_M}+\frac{Ze}{T}\partial_{\psi}\bar{\phi}_1-e^{-\psi/\psi_s}\int_{\psi_0}^{\psi}d\chi e^{\chi/\psi_s}\left(\frac{\partial^2_{\chi}F_M}{F_M}+\frac{Ze}{F_M}\partial_{\chi}\left(\frac{F_M}{T}\partial_{\chi}\bar{\phi}_1\right)\right)\right]\right\}-\\&-\frac{1}{f_t}\frac{1}{\{\nu_D^{ii}\}}\left\langle \frac{1}{hn}\int d^3v \nu_D^{ii}v_{\parallel}\left[F_M\frac{Ze\bar{\phi}_1}{T}-e^{-\psi/\psi_s}\int_{\psi_0}^{\psi}d\chi e^{\chi/\psi_s}Ze\partial_{\chi}\left(\frac{F_M}{T}\bar{\phi}_1\right)\right]\right\rangle
\label{soluz_per_ui}
\end{split}\end{equation}
The integrals in the last line of Eq.\ref{soluz_per_ui} can be computed, and the result is:
\begin{equation}\begin{split}
&\int d^3v \nu_D^{ii}v_{\parallel}F_M\frac{Ze\bar{\phi}_1}{T}=\frac{Ze\bar{\phi}_1}{T}\int d^3v \nu_D^{ii}v_{\parallel}F_M=0\\&
\int d^3v \nu_D^{ii}v_{\parallel}e^{-\psi/\psi_s}\int_{\psi_0}^{\psi}d\chi e^{\chi/\psi_s}Ze\partial_{\chi}\left(\frac{F_M}{T}\bar{\phi}_1\right)=e^{-\psi/\psi_s}\int_{\psi_0}^{\psi}d\chi e^{\chi/\psi_s}\partial_{\chi}\left(\frac{Ze\bar{\phi}_1}{T}\int d^3v \nu_D^{ii}v_{\parallel}F_M\right)=0
\end{split}\end{equation}
because $\nu_D^{ii}$ and $F_M$ are even functions of $v$, so that the argument of the integrals is odd. Using the solution Eq.\ref{soluz_per_ui} in Eq.\ref{f_1^0_completa}, we come to the solution:
\begin{equation}\begin{split}
&f_1^{(0)}=-\frac{Iv_{\parallel}}{\Omega}\left[\partial_{\psi}F_M+Ze\frac{F_M}{T}\partial_{\psi}\bar{\phi}_1-e^{-\psi/\psi_s}\int_{\psi_0}^{\psi}d\chi e^{\chi/\psi_s}\left(\partial^2_{\chi}F_M+Ze\partial_{\chi}\left(\frac{F_M}{T}\partial_{\chi}\bar{\phi}_1\right)\right)\right]+\\&+\frac{IHV_{\parallel}}{h\Omega}\left(\frac{mv^2}{2T}-1.33\right)\frac{d\log T}{d\psi}F_M-\frac{Ze\bar{\phi}_1}{T}F_M+e^{-\psi/\psi_s}\int_{\psi_0}^{\psi}d\chi e^{\chi/\psi_s}Ze\partial_{\chi}\left(\frac{F_M}{T}\bar{\phi}_1\right)-\\&-\frac{IHV_{\parallel}}{h\Omega}e^{-\psi/\psi_s}\int_{\psi_0}^{\psi} d\chi e^{\chi/\psi_s}\left[\partial^2_{\chi}F_M+Ze\partial_{\chi}\left(\frac{F_M}{T}\partial_{\chi}\bar{\phi}_1\right)-\frac{F_M}{\left\{\nu_D^{ii}\right\}}\left\{\nu_D^{ii}\left(\frac{\partial^2_{\chi}F_M}{F_M}+\frac{Ze}{F_M}\partial_{\chi}\left(\frac{F_M}{T}\partial_{\chi}\bar{\phi}_1\right)\right)\right\}\right]
\label{f1_short}
\end{split}\end{equation}

\section{Check of the drikt-kinetic limit}
In this short section I try again to recover the drift-kinetic solution in the limit of large wavelength. To do this, let's first consider an equation of this kind:
\begin{equation}
(1+\psi_s\partial_{\psi})f=K
\label{equaz_modello}
\end{equation}
In the limit $\psi_s\partial_{\psi}\ll1$ (equivalent to $k_{\perp}\rho_i\ll1$), Eq.\ref{equaz_modello} becomes $f=K$. The complete solution of Eq.\ref{equaz_modello} is
\begin{equation}
f=c_1e^{-\psi/\psi_s}+\frac{e^{-\psi/\psi_s}}{\psi_s}\int_{\psi_0}^{\psi}d\chi e^{\chi/\psi_s}K(\chi)
\label{soluz_modello}
\end{equation}
where $c_1e^{-\psi/\psi_s}$ is the homogeneous solution. In the limit $\psi_s\partial_{\psi}\ll1$, which implyes $\psi/\psi_s\gg1$, so that the homogeneous solution goes to zero. The particular equation can be integrated by parts:
\begin{equation}
\frac{e^{-\psi/\psi_s}}{\psi_s}\int_{\psi_0}^{\psi}d\chi e^{\chi/\psi_s}K(\chi)=K-\frac{e^{-\psi/\psi_s}}{\psi_s}\int_{\psi_0}^{\psi}d\chi e^{\chi/\psi_s}\psi_s\partial_{\chi}K(\chi)
\label{soluz_limite}
\end{equation}
but since $\psi_s\partial_{\psi}\ll1$, the second term in Eq.\ref{soluz_limite} is negligible and we are left with $K$. In the solution Eq.\ref{f1_short}, the arguments of the integrals contain derivatives respct to $\chi$, and so they are in the form Eq.\ref{soluz_limite}. Thus the limit of large wavelengths of Eq.\ref{f1_short} is
\begin{equation}
f_1^{(0)}=-\frac{Iv_{\parallel}}{\Omega}\frac{\partial F_M}{\partial \psi}+\frac{IHV_{\parallel}}{h\Omega}\left(\frac{mv^2}{2T}-1.33\right)\frac{d\log T}{d\psi}F_M-\frac{Ze\phi_1}{T}F_M
\label{f1_shorter}
\end{equation}
where I removed the bar symbol over $\phi_1$ because, in this limit, we can reasonably approximate the gyroaverage of $\phi_1$ with its value in the position of the guiding centers. Apart from the term proportional to the elctrostatic potential (which gives no contribution to the fluxes anyway), Eq.\ref{f1_shorter} corresponds to Eq.\ref{20}.

\section{Poloidal flow damping}
We use the flux-friction relation Eq.\ref{stress_friction} to compute the poloidal flow damping from Eq.\ref{f1_short}. First we have to find the parallel flow velocity $U_{\parallel i}$; to do that, we first have to expand the derivatives in Eq.\ref{f1_short}; in particular, the quantity $\partial^2_{\psi}F_M+Ze\partial_{\psi}(F_M/T_i\partial_{\psi}\bar{\phi}_1)$ becomes:
\begin{equation}\begin{split}
&F_M\left[\left(\frac{d\log P_i}{d\psi}\right)^2+\left(\frac{Ze}{T_i}\frac{d\bar{\phi}_1}{d\psi}\right)^2+\left(\frac{m^2v^4}{4T_i^2}-6\frac{mv^2}{2T_i}+\frac{25}{4}\right)\left(\frac{d\log T_i}{d\psi}\right)^2+3\frac{Ze}{T_i}\frac{d\bar{\phi}_1}{d\psi}\frac{d\log P_i}{d\psi}+\right.\\&\left.+2\left(\frac{mv^2}{2T}-\frac{5}{2}\right)\frac{d\log P_i}{d\psi}\frac{d\log T_i}{d\psi}+\left(3\frac{mv^2}{2T}-\frac{19}{2}\right)\frac{d\bar{\phi}_1}{d\psi}\frac{d\log T_i}{d\psi}+\frac{d^2\log P_i}{d\psi^2}+2\frac{Ze}{T_i}\frac{d^2\bar{\phi}_1}{d\psi^2}+\left(\frac{mv^2}{2T}-\frac{5}{2}\right)\frac{d^2\log T_i}{d\psi^2}\right]
\end{split}\end{equation}
Performing the calculations, we come up with this expression for the parallel velocity:
\begin{equation}\begin{split}
&U_{\parallel i}=-\frac{IT_i}{m_i\Omega_i}\left(\frac{d\log P_i}{d\psi}+2\frac{Ze}{T_i}\frac{d\bar{\phi}_1}{d\psi}-1.17\frac{d\log T_i}{d\psi}\right)+\frac{I}{\Omega_i}e^{-\psi/\psi_s}\int_{\psi_0}^{\psi}d\chi e^{\chi/\psi_s}\frac{T_i}{m_i}\left[\left(\frac{d\log P_i}{d\psi}\right)^2+2\left(\frac{Ze}{T_i}\frac{d\bar{\phi}_1}{d\psi}\right)^2-\right.\\&\left.-\left(\frac{d\log T_i}{d\psi}\right)^2+\frac{d\log P_i}{d\psi}\left(3\frac{Ze}{T_i}\frac{d\bar{\phi}_1}{d\psi}-2\frac{d\log T_i}{d\psi}\right)-5\frac{Ze}{T_i}\frac{d\log T_i}{d\psi}\frac{d\bar{\phi}_1}{d\psi}+\frac{d^2\log P_i}{d\psi^2}+2\frac{Ze}{T_i}\frac{d^2\bar{\phi}_1}{d\psi^2}-\frac{d^2\log T_i}{d\psi^2}\right]-\\&-\frac{I}{\Omega}e^{-\psi/\psi_s}\int_{\psi_0}^{\psi}d\chi e^{\chi/\psi_s}\frac{T_i}{m_i}\left[0.35\frac{d\log P_i}{d\psi}\frac{d\log T_i}{d\psi}+0.52\frac{Ze}{T_i}\frac{d\bar{\phi}_1}{d\psi}\frac{d\log T_i}{d\psi}+0.17\frac{d^2\log T_i}{d\psi^2}-2.27\left(\frac{d\log T_i}{d\psi}\right)^2\right]
\label{parallel_flow_long}
\end{split}\end{equation}
The coefficients in the last line come from the formulas for the velocity averages reported at the end of the book "Collisional transport in magnetized plasmas" by Helander \& Sigmar. Using the solution Eq.\ref{f1_short} in Eq.\ref{formula_85} and using the result Eq.\ref{parallel_flow_long}, we arrive at the following expression (which is acted on by the operator $e^{-\psi/\psi_s}\int d\chi e^{\chi/\pi_s}$):
\begin{equation}\begin{split}
&\left\langle B\frac{IT_i}{m_i\Omega_i}\int mv_{\parallel}\frac{hv_{\parallel}-HV_{\parallel}}{hT_i}F_M\nu_D^{ii}\left[\left(x^4-6x^2+\frac{25}{4}\right)\left(\frac{d\log T_i}{d\psi}\right)^2+\left(2x^2-5\right)\frac{d\log T_i}{d\psi}\frac{d\log P_i}{d\psi}+\right.\right.\\&\left.\left.+\left(3x^2-\frac{19}{2}\right)\frac{Ze}{T_i}\frac{d\bar{\phi}_1}{d\psi}\frac{d\log T_i}{d\psi}+\left(x^2-\frac{5}{2}\right)\frac{d^2\log T_i}{d\psi^2}\right]\right\rangle+\left\langle B\frac{IT_i}{m_i\Omega_i}\int mv_{\parallel} \frac{HV_{\parallel}}{hT_i}F_M\nu_D^{ii}\left[-1.27\left(\frac{d\log T_i}{d\psi}\right)^2+\right.\right.\\&\left.\left.+2.35\frac{d\log T_i}{d\psi}\frac{d\log P_i}{d\psi}+5.52\frac{Ze}{T_i}\frac{d\bar{\phi}_1}{d\psi}\frac{d\log T_i}{d\psi}+1.17\frac{d^2\log T_i}{d\psi^2}\right]\right\rangle+\left\langle B\frac{IT_i}{m_i\Omega_i}\int mv_{\parallel}^2F_M\nu_D^{ii}\left[3.27\left(\frac{d\log T_i}{d\psi}\right)^2+\right.\right.\\&\left.\left.+1.65\frac{d\log T_i}{d\psi}\frac{d\log P_i}{d\psi}+4.48\frac{Ze}{T_i}\frac{d\bar{\phi}_1}{d\psi}\frac{d\log T_i}{d\psi}+0.63\frac{d^2\log T_i}{d\psi^2}\right]\right\rangle
\label{flow_damping_lung}
\end{split}\end{equation}
where $x^2=mv^2/(2T)$. The numerical coefficients come from the averages over the velocity and the flux surface. Eq.\ref{flow_damping_lung} consists in the contribution to the poloidal flow damping from the FLR effects. Performing also the last integrals and adding the result from the large-wavelength part of the distribution function: 
\begin{equation}\begin{split}
\left\langle \boldsymbol B\cdot\nabla\cdot\boldsymbol \pi_i\right\rangle\approx B\mu_{01}m_in\nu_{ii}\frac{IT_i}{m_i\Omega_i}&\left\{1.17\frac{d\log T_i}{d\psi}+e^{-\psi/\psi_s}\int_{\psi_0}^{\psi}d\chi e^{\chi/\psi_s}\left[-4.54\left(\frac{d\log T_i}{d\psi}\right)^2+\right.\right.\\&\left.\left.+0.70\frac{d\log T_i}{d\psi}\frac{d\log P_i}{d\psi}+1.04\frac{Ze}{T_i}\frac{d\bar{\phi}_1}{d\psi}\frac{d\log T_i}{d\psi}+0.54\frac{d^2\log T_i}{d\psi^2}\right]\right\}
\label{flow_damping_def}
\end{split}\end{equation}
where $\mu_{01}=\{\nu_D^{ii}\}\approx0.53$. The first term proportional to $d\log T_1/d\psi$ comes from the drift-kinetic theory. The coefficients multiplying the other terms were obtained by using Eq.\ref{19+1} and Eq.\ref{19+2}. Apart from the numerical coefficients, we can see that new contributions appear from the gradients of temperature, pressure and the electrostatic potential. The first two terms in the square brackets can be written also as
\begin{equation}\begin{split}
-4.54\left(\frac{d\log T_i}{d\psi}\right)^2+0.70\frac{d\log T_i}{d\psi}\frac{d\log P_i}{d\psi}=-3.84\left(\frac{d\log T_i}{d\psi}\right)^2+0.70\frac{d\log T_i}{d\psi}\frac{d\log n_i}{d\psi}
\label{flow_damping_alt}
\end{split}\end{equation}
where I used $P_i=n_iT_i$ to make explicit the dependence from the density, which is more useful for an eventual use of this result in a four-field gyrofluid model. The expression Eq.\ref{flow_damping_def} is still in an integral form because it contains the FLR corrections. Since this expression has been obtained by solving formally the gyrokinetic equation, although under specific hypotheses, it contains the corrections to all orders in the FLR parameter, which can be written in the form $k_{\perp}\rho_i$, where $k_{\perp}$ is the inverse of the typical length-scale of the gradients. In the drift-kinetic limit $k_{\perp}\rho_i\ll1$, thus in this case an expansion in powers of this parameter makes sense; in the gyrokinetic case, instead, the condition $k_{\perp}\rho_i=O(1)$ holds, so that the power expansion results inadequate. For this reason, a solution such as Eq.\ref{flow_damping_def} is suitable for being included in a system of gyrofluid equations.

\end{document}